# Excitation and manipulation of super cavity solitons in multi-stable passive Kerr resonators

Pengxiang Wang[1], Jianxing Pan[2], Tianye Huang[2,3,4], Shengbo Xu[1], Ran Xia[1], Julien Fatome[5], Bertrand Kibler[5], Carlos Mas-Arabi[6] and Gang Xu[1]

[1] School of Electronic information, Huazhong University of Science and Technology, Wuhan, China
[2] School of Mechanical Engineering and Electronic, China University of Geosciences, Wuhan, China
[3] Wuhan National Laboratory for Optoelectronics, Huazhong University of Science and Technology, Wuhan, China
[4] Shenzhen Research Institute of China University of Geosciences, Shenzhen, 518000, China
[5] Laboratoire Interdisciplinaire Carnot de Bourgogne (ICB), UMR 6303, Université de Bourgogne, 9 Avenue Alain Savary, 21078, Dijon, France
[6] Institut Universitari de Matematica Pura i Aplicada, Universitat Politecnica de Valencia, Valencia, Spain

**Abstract**

We report on the theoretical analysis as well as the numerical simulations about the nonlinear dynamics of cavity solitons in a passive Kerr resonator operating in the multistable regime under the condition of a sufficiently strong pump. In this regime, the adjacent tilted cavity resonances might overlap, thus leading to the co-existence of combinatory states of temporal cavity solitons and the extended modulation instability patterns. Very interestingly, the cavity in the regime of multistablity may sustain distinct families of cavity solitons, vividly termed as *super cavity solitons* with much higher intensity and broader spectra if compared with those in the conventional bi-stable regime. The description of such complex cavity dynamics in the multstable regime requires either the infinite-dimensional Ikeda map, or the derived mean-field coupled Lugiato-Lefever equations by involving the contributing cavity resonances. With the latter model, for the first time, we revealed the existence of different orders of super cavity solitons, whose stationary solutions were obtained by using the Newton-Raphson algorithm. Along this line, with the continuation calculation, we have plotted the Hopf / saddle-node bifurcation curves, thus identifying the existing map of the stable and breathing (super) cavity solitons. With this defined parameter space, we have proposed an efficient method to excite and switch the super cavity solitons by adding an appropriate intensity (or phase) perturbation on the pump. Such deterministic cavity soliton manipulation technique is demonstrated to underpin the multi-level coding, which may enable the large capacity all-optical buffering based on the passive fiber ring cavities.

## 1. Introduction

Kerr nonlinearly driven cavities exhibit a rich variety of physical phenomena such as cavity solitons (CSs) [1],[2], frequency combs [3], modulation instabilities [4], self-pulsing, bistability, chaos [5],[6],[7] and so on. These interesting pumping-dissipating complex dynamics have been successfully observed experimentally in passive Kerr resonators, including fiber rings, Fabry-Perot (FP) resonators and integrated microresonators [8],[9]. In general, the Ikeda-map [10], i.e, the nonlinear Schrödinger equation combined with the discrete cavity boundary conditions, provides a straightforward numerical model to describe the optical field evolution. With the good cavity approximation (low insertion loss), the simple but accurate mean-field Lugiato-Lefever equation (LLE) [11] was derived as a powerful tool to simplify the analysis of nonlinear propagation inside the cavity, not only for the continuous-wave (CW) or the homogeneous steady states (HSS) solutions of this model [12] but also for the localized structures [13]. For the synchronously driven cavities with anomalous dispersion operating in the bistable state, the nonlinear coherent resonances might be tilted and spread to a "soliton step" [14],[15], where the CSs could be excited either by scanning the cavity detuning across the regime of modulation instability (MI) or via the mechanical perturbations [16],[17]. More interestingly, in recent years, the coupling dynamics of CSs [18], namely the generation of soliton molecules [19] and vector CSs [20],[21] in two orthogonal polarization states have been observed in passive cavities. These soliton synthesis mechanisms could be mainly attributed to the cross-phase modulation (XPM) among different components [16], and the relevant binding evolutions have been satisfactorily described by the vector form of LLE [18],[20],[21].

However, when the pump power exceeds certain levels, the tilted cavity resonance might eventually overlap with the adjacent one. In this case, the cavity enters into the multi-stable states [22]. Here, systematic descriptions of the nonlinear dynamics are beyond the reach of the aforementioned form of the LLE solely adequate for a single resonance, but this mysterious regime is indeed attractive due to the various hidden combinatory nonlinear states, including MI patterns, CSs and so on [23],[24]. In this framework, in 2015, Hansson and Wabnitz predicted that with the joint contribution of multiple cavity resonances, a new family of CSs could be excited with much higher intensities and broader spectra [22]. These specific CSs might be termed as *super cavity solitons* (SCSs), whose traces have been recorded by Anderson *et al.* in a passive fiber ring by scanning the cavity detuning [25]. However, the experimental results in ref. [25] were not clearly understood due to the lack of available analytical treatments. Despite several attempts to derive the mean-field equations to involve multiple resonances, namely the coupled Lugiato-Lefever equations (CLLEs) [26],[27], the nonlinear dynamics of SCSs are not completely characterized. Until now, efficient excitation and manipulation techniques for SCSs are still pending because the parameter space and the relevent stability analysis for different families of CSs / SCSs are yet to be explored.

In this work, for the first time, we provide the full parameter space for the CSs and SCSs based on the CLLEs by essentially involving the principal contributing cavity resonances. With the Newton-Raphson method and the numerical continuation, we plotted the existing map of SCSs where the stable and breathing/chaotic SCSs are successfully identified with the relevant stability analysis in the function of two key driving parameters, i.e., the pump power and the cavity detuning. More specifically, we have uncovered the coexistence of distinct orders of SCSs sustained in the multistable states of the cavity, where adequate phase/intensity perturbations might enable the deterministic switching and manipulation of CS and distinct orders of SCSs. Based on these salient properties, we performed the efficient ASCII coding of arbitrary letters with the multi-level solitons in a fiber ring with moderate parameters. Along this line, the writing, modifying, and erasing of these multi-level pulses are demonstrated, thus opening the way for high-capacity optical telecommunications and all-optical buffers based on the SCSs.

## 2. Methods

### 2.1. Ikeda-map model

In order to clearly uncover the emergent mechanism of SCSs, it is essential to start with the Ikeda-map, which is a universal and accurate approach [10],[28],[29],[30] to characterize the nonlinear cavity dynamics involving single or multiple resonances. In short, this model contains a periodic boundary condition and the nonlinear Schrödinger equation (NLSE). It can be written in the following form:

$$\begin{cases} \frac{\partial E^m(t,z)}{\partial z} = -\frac{\alpha_i}{2} E^m(t,z) - i\frac{\beta_2}{2}\frac{\partial^2 E^m(t,z)}{\partial t^2} + i\gamma |E^m(t,z)|^2 E^m(t,z) \\ E^{m+1}(t,0) = \sqrt{\theta} E_{in} + \sqrt{1-\theta} e^{i\delta_0} E^m(t,L). \end{cases} \quad (1)$$

In these equations, $z$ denotes the propagation distance along the circular resonant cavity, and $t$ describes the fast time of co-rotation with the intracavity field. $\theta$ is the power coupling coefficient and $E_{in}$ is the pump field, where the subscript $m$ describes the roundtrip number, and $\delta_0$ the roundtrip phase detuning between the pump field and the closest resonance. Here, $\beta_2$ is the group velocity dispersion (GVD) coefficient, $\gamma$ is the nonlinear coefficient, and $\alpha_i$ is the linear loss coefficient.

For Eq. (1), a special solution is the CW steady state ($\partial E^m(z,\tau)/\partial \tau = 0$). In this case, the dependence of the input power $P_{in} = |E_{in}|^2$ and the intra-cavity power $P = |E^m|^2$ satisfies the Airy equation $P = \theta P_{in}/\{(1-\sqrt{\rho})^2 [1 + F sin^2(\frac{\delta_0 - \gamma LP}{2})]\}$. This equation is well-known to depict the variation of the cavity power in the FP resonator, where a series of periodic tilted resonances could emerge by scanning the cavity detuning $\delta_0$. For a considerable level of the pump power, these resonances become titled due to the nonlinear phase shift induced by the self-phase modulation (SPM). When the maximum of this phase shift is larger than the resonance width, the CW response of the system becomes multivalued and enter into bistability regime. Eventually, as the amount of nonlinear detuning continues to increase, adjacent resonances gradually overlap with each other. In addition to the bistability associated with a single resonance, such a large nonlinear resonance tilting induces the CW tristability which means that the system has three homogeneous equilibrium points that are stabilized under the CW state. In this work, the optical fiber ring architecture under investigation is depicted in Fig. 1(a) and the key experimental parameters (quoted in the figure captions) are similar to those in the Ref. [25] where the nonlinear tilted-cavity resonances overlap each other. In this particular regime rarely explored, the conventional LLE model, which consider only a single resonance, thus failing to provide sufficient physical insights for the tri-stability or multi-stability operation.

## 2.2. CLLEs model

Due to the shortcoming of the LLE mode, Conforti and Biancalana derived the coupled Lugiato-Lefever equations (CLLE) based on the Ikeda-map (1) [26]. Its general form could be written as:

$$i\frac{\partial U_n}{\partial z} - \frac{\delta_n}{L}U_n - \frac{\beta_2}{2}\frac{\partial^2 U_n}{\partial t^2} + \gamma \sum_{p=-N_R}^{N_L} \sum_{q=q_{min}}^{q_{max}} U_p U_q U_{p-n-q}^* = i\frac{\sqrt{\theta}}{L}E_{in} - i\frac{\alpha_0}{L}\sum_{p=-N_R}^{N_L} U_p \quad (n = -N_R \cdots N_L) \quad (2)$$

where $U_n = E_n \exp\left(\frac{i\delta_0 z}{L}\right)$, $q_{min} = \max\{-N_R, n-p-N_R\}$, $q_{max} = \min\{N_L, n-p+N_L\}$, $\delta_n = \delta_0 + 2\pi n$. The half fraction of power lost per roundtrip is $\alpha_0 = (\alpha_i L + \theta)/2$. In this equation $U$ is the intra-cavity field and its subscript $n$ labels the number of the involved resonances. The two key driving parameters, i.e. the cavity detuning and the external pump are denoted by $\delta_n$ and $E_{in}$, respectively. Very interestingly, the fourth term on the LHS takes into account that not only the SPM but also the XPM of distinct resonances and each of them is written by an LLE-type equation that are coupled in a nontrivial way. In other words, the CLLE model (2) refines the inability of the mean-field model to describe the excessive overlap of resonance peaks. Therefore, it is capable to describe the nonlinear cavity dynamics operating in the multi-stable regime. In fact, one may adopt the simplified form of Eq. (2), e.g. the degenerated conventional LLE for $n = 1$, which is sufficiently accurate in most cases. While in this work to investigate the multi-stable cavities, it is essential to take into account at least 3 interactive resonances and this model is simplified to be the third-order form ($N_L = 2$, $N_R = 0$) and the governing model could be written as the reduced 3-CLLE system:

$$\begin{cases} i\frac{\partial U_0}{\partial z} - \frac{\delta_0}{L}U_0 - \frac{\beta_2}{2}\frac{\partial^2 U_0}{\partial t^2} + \gamma(|U_0|^2 + 2|U_1|^2 + 2|U_2|^2)U_0 + \gamma U_1^2 U_2^* \\ \qquad = i\frac{\sqrt{\theta}}{L}E_{in} - i\frac{\alpha_0}{L}(U_0 + U_1 + U_2) \\ i\frac{\partial U_1}{\partial z} - \frac{\delta_0+2\pi}{L}U_1 - \frac{\beta_2}{2}\frac{\partial^2 U_1}{\partial t^2} + \gamma(|U_1|^2 + 2|U_0|^2 + 2|U_2|^2)U_1 + 2\gamma U_0 U_1^* U_2 \\ \qquad = i\frac{\sqrt{\theta}}{L}E_{in} - i\frac{\alpha_0}{L}(U_0 + U_1 + U_2) \\ i\frac{\partial U_2}{\partial z} - \frac{\delta_0+4\pi}{L}U_2 - \frac{\beta_2}{2}\frac{\partial^2 U_2}{\partial t^2} + \gamma(|U_2|^2 + 2|U_0|^2 + 2|U_1|^2)U_2 + \gamma U_1^2 U_0^* \\ \qquad = i\frac{\sqrt{\theta}}{L}E_{in} - i\frac{\alpha_0}{L}(U_0 + U_1 + U_2). \end{cases} \quad (3)$$

In these coupled equations, $U_0$ is the optical field originating from the principal resonance, while $U_1$ and $U_2$ are those based on the two additional negatively detuned resonances. From these equations, one may clearly get the physical insight of the contributing components to the nonlinear evolution of $U_n$ and they provide reliable bases to perform the stability analysis of CSs and SCSs.

## 2.3. Stability analysis and continuation

Based on the 3-CLLE (3) in the previous section, we apply the linear stability analysis [31],[32] to the cavity operating in the multi-stable regime. Particularly, we can obtain more accurate CS and SCS solutions of the 3-CLLE (3) by adopting the Newton-Raphson algorithm [35], which have been widely used in the scalar LLE to analyze various nonlinear dynamics, including the Hopf bifurcations, the symmetry breaking, the switching waves [26],[33],[34] In this way, we can obtain the coordinate parameters of the CS and SCS from the Ikeda-map model, which are brought into the CLLE model as the initial condition for numerical continuation. Moreover, the calculation of the Jacobian matrix $J(X_0)$ eigenvalues is essential to analyze the stability of the system. We refer the interested readers to the appendix section for more details about the mathematical routines. In Fig. 1(b), we plotted the soliton branches based on the above-mentioned numerical continuation associated with the stability analysis of the 3-CLLE (3), where the solid-dark lines indicate the peak power of stable CSs or the SCSs. The soliton branches refer to the peak power of the stable or unstable solitons, namely the breathing or chaotic solitons. The blue, green and red dots in these curves are the saddle-node (SN) bifurcation points for CSs and SCSs, whose detailed mathematical properties are revealed in the appendix section. It is worth mentioning that with this level of pump power, we discovered the coexistence of CSs and different orders of SCSs for a fixed normalized cavity detuning at $\delta_1 = 4$. Here, the 3 intersection points with the soliton branches correspond to the soliton peak power. To be more specific, in Fig. 1(c), we plotted the intensity profiles of the CS, the $SCS_1$ (1st order) and the $SCS_2$ (2nd order) based on the Newton Raphson algorithm. One may also notice that the SCSs (especially those with higher order) have higher peak power and shorter pulse width. More salient properties of SCSs would be revealed in the next sections.

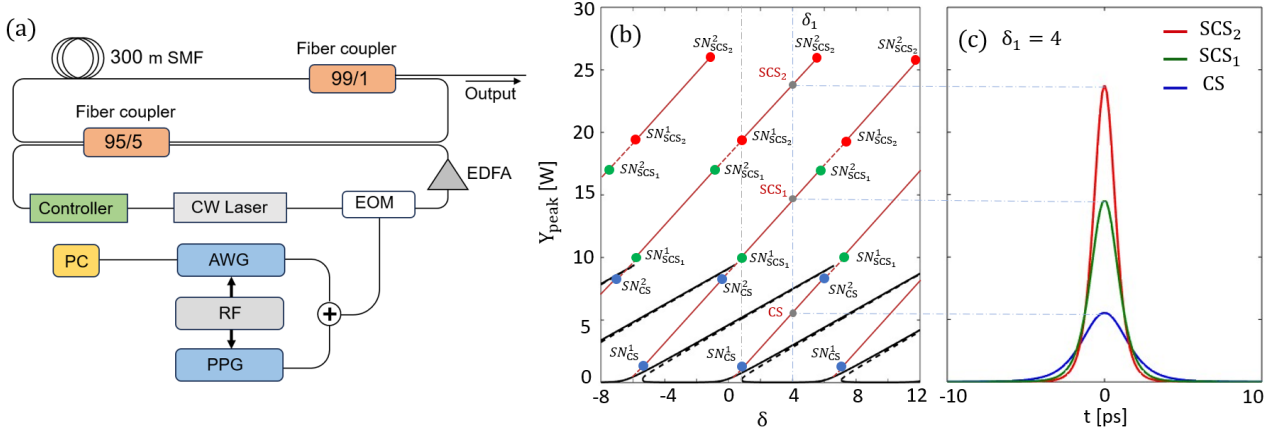

FIG. 1 **Multistable dissipative fiber ring and localized structures in each steady state.** (a) Corresponding typical experimental configuration analyzed. AWG: arbitrary waveform generator, EOM: electro-optic modulators, EDFA: erbium-doped optical fiber amplifier, PC: personal computer, PPG: pulse pattern generator, RF: radio frequency. The cavity is made with standard single-mode fiber (SMF-28), whose resonator roundtrip length is $L = 300$ m, the group velocity dispersion is $\beta_2 = -22.9$ ps$^2$/km and the nonlinear coefficient $\gamma = 4.6$ (W·km)$^{-1}$ for the pump wavelength at ~ 1550 nm with the peak power $P_0 = 0.8$ W. Fiber couplers (input loss $\theta_1 = 0.05$, output loss $\theta_2 = 0.01$, total coupling loss $\theta = \theta_1 + \theta_2$) are used to incorporate the total round-trip loss into the parameter $\alpha_i = 0.2$ km$^{-1}$, $\alpha_0 = 0.06$, so we have the cavity finesse $\mathcal{F} \sim 52$. (b) Cavity resonances (dark lines) and soliton branches (red lines). The solid red lines refer to the soliton states (stable or unstable). Blue, green and red dots are the saddle-node bifurcation points for CS, SCS$_1$ and SCS$_2$, respectively. (c) Theoretical calculation of cavity soliton intensity profiles via the Newton-Raphson algorithm. With the fixed cavity detuning $\delta_1 = 4$, CS (blue curve), SCS$_1$ (green curve) and SCS$_2$ (red curve) are simultaneously excited from the three soliton branches shown in (b).

## 3. Results and discussion

3.1. Excitation and the switching dynamics of cavity solitons

In the previous section, we have demonstrated the co-existence of different families of CSs and SCSs under the same cavity driving condition with sufficiently tilted cavity resonances. However, the direct excitation of SCSs is not a trivial task, even though their traces have been recorded by scanning the cavity detuning in ref. [25], where the generation of stable SCSs is still lacking. Indeed, we have tested the method of detuning scan in the numerical simulations of Ikeda map (1), where the SCSs and is usually annihilated due to the perturbation suffered when δ is swept across the unstable soliton regimes [shown in Fig. 1(b)] of distinct resonances. Therefore, a reliable excitation method of SCSs is highly desired for the current study. Fortunately, the generation of CSs is guaranteed by scanning and then locking of the detuning and we have discovered a reproducible routine to switch the CSs to SCSs by adding appropriate strength of phase or intensity perturbations. As shown in Fig. 2(a), we may write a CS from the CW by adding a Gaussian-form signal to the soliton spot on the pump at the $m_0^{th}$ roundtrip, which can be written as $E_{in,m}(\tau) = E_{in} + \sqrt{P_P}e^{-t^2/\Delta t^2}\delta(m - m_0)$. This method could be realized in experiments by superposing an addressing beam on the CW pump [28],[36],[37],[38]. Furthermore, we have been inspired by the all-optical manipulation experiments of the co-existing polarization states of the CSs by adding a slight dipping or a bumping perturbation to trigger the deterministic switching of their polarization states [20]. Here, we implemented this approach by adding the bump for a single roundtrip and successfully led to the transition from the CS to the SCS$_1$ and then the SCS$_1$ to SCS$_2$. Similarly, adding the dipping perturbations on the pump could also switch back the SCS$_2$ to SCS$_1$ and then the SCS$_1$ to the CS. As in the ref. [36],[37],[38], one may finish the full switching cycle by erasing the CS with this method. The intensity and the spectra evolution of CS, SCS$_1$, and SCS$_2$ are shown in Fig. 2(a) and 2(b), respectively. In these 2 subfigures, it is obvious that the SCS$_2$ and SCS$_1$ have much higher peak power and much broader spectra.

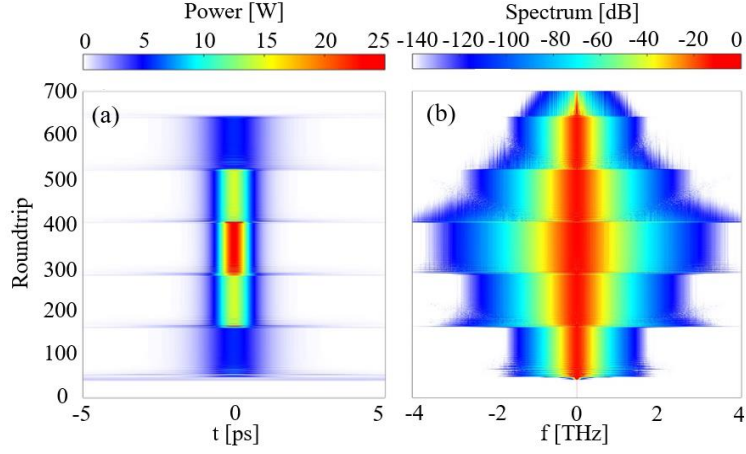

FIG. 2 **Switching dynamics between distinct families of CSs**. Soliton traces in the temporal (a) and spectral (b) domain are obtained from the numerical simulations based on the Ikeda map (1), where the key parameters are similar to those in Fig. 1. The one-roundtrip intensity perturbations on the pump $P_P$ are roughly selected with 23 W, 11 W, 11 W, 7 W, 7 W and 5W at the 40$^{th}$, the 160$^{th}$, the 280$^{th}$, the 400$^{th}$, the 520$^{th}$ and the 640$^{th}$ roundtrip, respectively.

3.2. Parameter space of super cavity solitons

In Fig. 2(a) we may notice that with the pump power of $P = 0.8$ W and the cavity detuning of $\delta = 4$, CSs, SCS$_1$ and SCS$_2$ could co-exist and stably propagate. One may wonder if there are certain ranges of parameters to target the stable CSs or the SCSs (or stable broadband frequency comb) that could be stabilized and circulate continuously inside the cavity. From this point of view, in this section, we systematically search for their normalized parameter space ($\Delta$, $S$) using numerical continuation and stability analysis [6],[39]. To keep the generality of the parameter space, we use the key parameters in the normalized form: cavity detuning $\Delta = \delta_0/\alpha_0$, pumping $S = \sqrt{\theta_1 \gamma L/\alpha_0^3}\, E_{in}$ and the fast time $\tau = \sqrt{|\beta_2|L/2\alpha_0}\, t$. To implement the Newton Raphson algorithm, we can write the initial of the solitons' complex amplitude in the following form: $E^{(0)}(z=0) = \sqrt{2(\delta_0 + 2n\pi)/\gamma L}\, \mathrm{sech}(\sqrt{2(\delta_0 + 2n\pi)/|\beta_2|L}\, t)$, where n = 0, 1, 2 correspond to CS, SCS$_1$ and SCS$_2$, respectively. In this way, we obtained the stationary solutions for CSs and SCSs with certain values of $\Delta$ and $S$. Then, we search for the bifurcation curves, which divide the parameter space into distinct characteristic regimes [40]. In Fig. 3, the frontier lines of distinct areas are mainly calculated by analyzing the real and the imaginary parts of the Jacobian matrix $J(X_0)$ of the 3-CLLE (3). To test these frontier curves, we have performed the numerical simulations based on the Ikeda map to confirm the stabilities. For example, we start from the original point ($S = 18$, $\Delta = 60$), when the system crosses the Hopf bifurcation curve, CS, SCS$_1$ and SCS$_2$ are no longer stable and displays time-period oscillations. On the bottom of the parameter space in Fig. 3(a-c), we may obtain the first saddle-node bifurcation curve ($SN^1$). In the area below this curve, neither CSs nor SCSs could survive and they will vanish and degenerate to the HSS. While on top of the parameter space, there exists the second saddle-node bifurcation curve ($SN^2$). In the area above this regime, there is no background to support the solitons and the MI patterns may appear. We should mention that there also exists the fourth characteristic line, i.e. the Belyakov Devaney curve (BD) and beyond this curve, the solitons could no longer be localized on the fast time. Instead, the spatiotemporal chaos of solitons could be observed in the numerical simulations [40]. From Fig. 3(a-c), we notice that the CS, SCS$_1$ and the SCS$_2$ exhibit similar stability features determined by the 4 aforementioned characteristic lines.

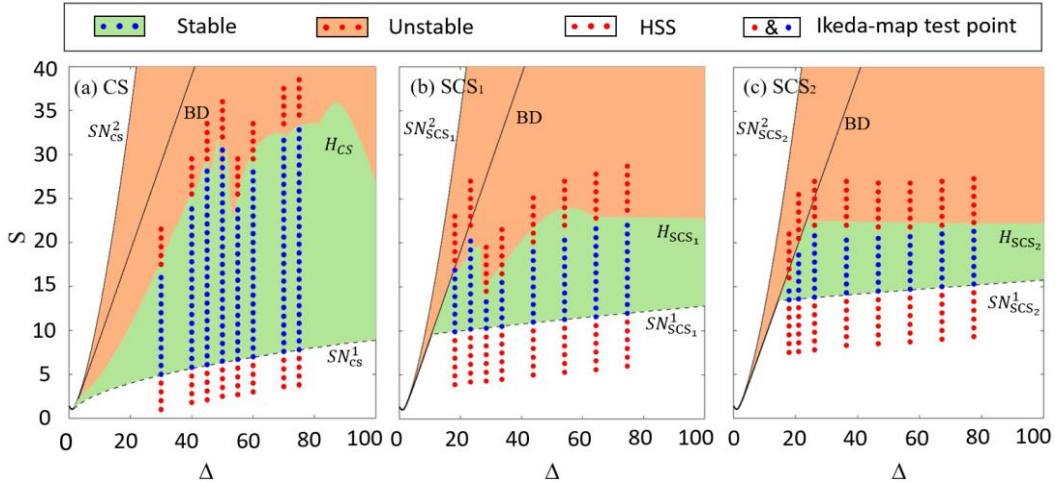

FIG.3 **Parameter space (Δ, S) with the stability information for CS (a), SCS$_1$ (b) and SCS$_2$ (c).** The light green areas represent stable CSs or SCSs regimes, while in the orange areas, these localized structures experience the regular or chaotic oscillating on the intensities. In white areas with large pump power or large cavity detuning, CSs or SCSs are unable to survive and they collapse to the HSS. Abbreviations in the figure, BD: Belyakov-Devaney line; H: Hopf bifurcation curves; $SN^1$ and $SN^2$: the 1st and the 2nd saddle-node bifurcation curves. Blue and red dots indicate the numerical tests based on the Ikeda map model (1) to confirm the temporal stability of the system.

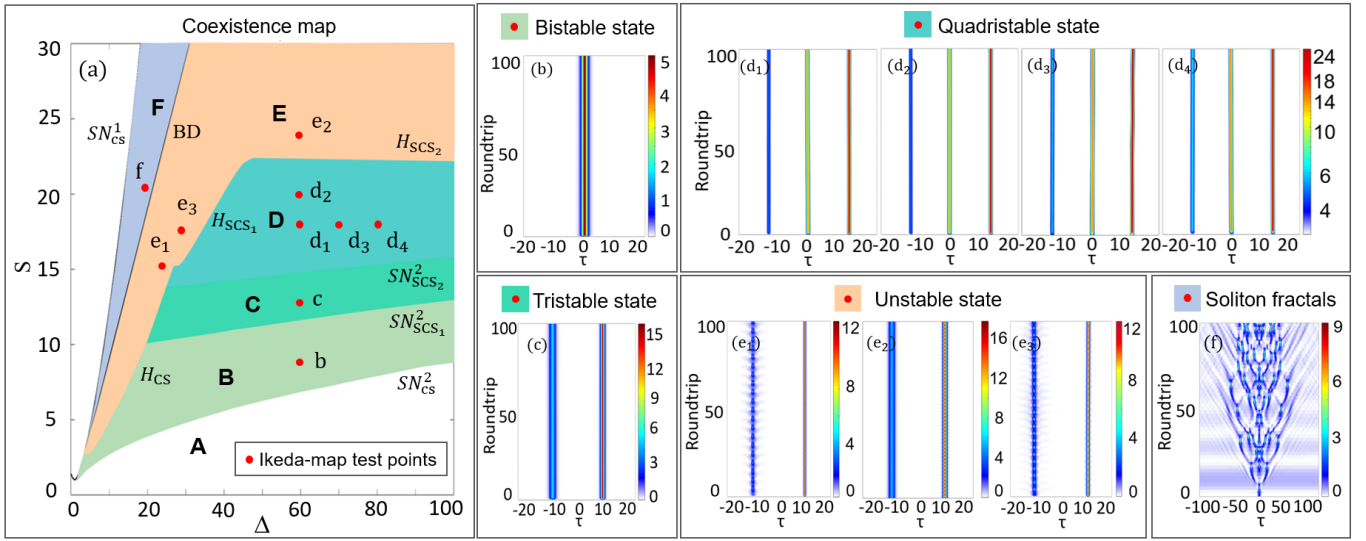

FIG. 4 **Coexistence landscape of CS, SCS$_1$ and SCS$_2$ in the parameter space (Δ, S).** (a) Combinatory states for the passive resonator sustaining multi-stable CSs. Distinct areas are colored white (only HSS, A), lime green (bistable state, B), mint green (tristable state, C), forest green (quadristable state, D), brown (unstable state, E) and light blue (soliton fractals, F). Their frontiers are the characteristic lines obtained by calculating the eigenvalues of the Jacobian matrix of the 3CLLE. Red dots in this figure indicate the location of the test points based on the Ikeda map model simulation and the corresponding intensity evolution of the solitons are demonstrated in (b-e). (b) Conventional CS in the bistable regime for (Δ, S) = (60,8). (c) CS and SCS$_1$ in the tristable regime for (Δ, S) = (60,12). (d) CS, SCS$_1$ and SCS$_2$ in the quadristable regime. In this area, we have tested 4 different positions at $d_1(\Delta, S) = (60, 18), d_2(\Delta, S) = (60, 20), d_3(\Delta, S) = (70, 18)$ and $d_4(\Delta, S) = (80, 18)$. One may notice that in this regime, the minor increment of $S$ lead to the negligible change on the soliton intensity (see $d_1 \to d_2$), while this changement could be obviours by slightly modifying the cavity detuning δ (see $d_1 \to d_3 \to d_4$). (e) The evolution of different families of oscillating solitons in unstable regime of the parameter space. We select three representative cases at (Δ, S) = (25, 15) with an Oscillating CS + a stable SCS$_1$ in (e1), an oscillating SCS$_1$ + a stable CS at (Δ, S) = (60, 24) at (e2), an oscillating CS + an oscillating SCS$_1$ at (Δ, S) = (28, 17.5) in (e3). (f) The spatiotemporally unstable solitons that may experience the fractal and chaos during the evolution. A representative case at (Δ, S) = (18, 20.5) is demonstrated in (f).

For higher-order stable SCSs, their existing parameter space tend to be narrower, and it isn't trivial to determine their coexistence area due to the overlap of the cavity resonances. Therefore, it is essential to superpose their parameter space to obtain more physical insights into such a highly multi-stable landscape of the cavity dynamics. As shown in Fig. 4(a), the soliton-existing regimes are painted with different colors. In the green areas (marked by B, C and D), stable CSs or SCSs may survive without intensity oscillations. We have selected several representative positions and performed the test with the Ikeda-map based numerical simulations to investigate the soliton propagation. As shown in Fig. 4(b-d), in the bi-stable, tri-stable and quadri-stable states, one may obtain the combinatory state of CS, CS+$SCS_1$ and CS+$SCS_1$+$SCS_2$, respectively. In the orange area, these solitons cross the Hopf bifurcation curves, leading to the intensity oscillations on the slow time, but still localized on the fast time [shown in Fig. 4(e1-e3)]. However, in the light blue area, these solitons may experience the spatiotemporal instabilities, and they are no longer localized in the fast time. As demonstrated in Fig. 4(f), we have tested these properties by setting CSs or SCSs as the initial condition, which could exhibit the fractal-like behaviors and eventually evolve to soliton chaos.

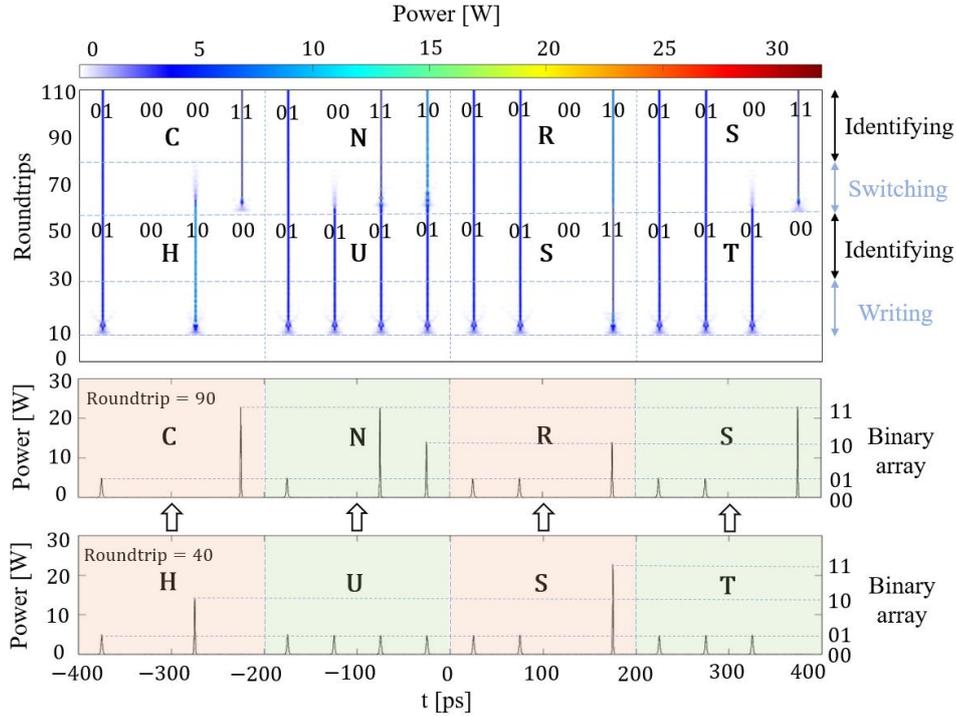

**FIG.5 Demonstration of the ASCII coding and the all-optical buffer of arbitrary letters by deterministic manipulation of SCSs.** In the 100 roundtrips, we complete the whole processing of the writing, identification, switching and re-identification in the Ikeda-map based simulation. The coding examples are the abbreviation for our joint team's affiliations, i.e. Huazhong University of Science and Technology (HUST) in China and the Centre National de la Recherche Scientifique (CNRS) in France. Key cavity parameters are similar to those in Fig. 1-2.

In the end, we briefly comment on the proof-of-concept applications of the high-capacity bit coding and the all-optical buffer based on the deterministic switching of CSs and SCSs. As shown in Fig. 4(a), one may select an arbitrary location in the quadristable regime where the CS, the $SCS_1$, and the $SCS_2$ can coexist. This 4-level-soliton configuration could be profited to implement the efficient binary coding, where the HSS, CS, $SCS_1$ and $SCS_2$ could be coded as 00, 01, 10 and 11 respectively. With this idea in our simulations, the time slot for each soliton is set to be 50 ps. Each can store 2 bits without any risk of the physical crosstalk if compared with the conventional 2-level soliton configuration. Therefore, 4 solitons are sufficient to constitute an 8-bit array ASCII code as an arbitrary letter or a symbol. Along this line, in Fig. 5 we demonstrate the potential application of this technique. At first, at the 10[th] roundtrip, different levels of intensity perturbations are added at the 16 soliton spots to write the four letters HUST [shown in Fig. 5], which is the abbreviation of the authors' research affiliation Huazhong University of Science and Technology in China. From the experimental point of view, as shown in Fig. 1(a), this process could be realized with the amplified signal from the AWG, which is synchronized with the quasi-CW pump. Therefore, the perturbation signal could be perfectly located at the desired soliton spots.

Theoretically, due to the fact that the three families of solitons are all stable under these cavity parameters, the signal of HUST could circulate endlessly inside the cavity. Moreover, at the 60[th] roundtrip, programmed perturbation signals are added to switch the signal from HUST to CNRS, which is the abbreviation of the authors' research affiliation Centre National de la

Recherche Scientifique in France. In this way, this cavity configuration with the presence of SCSs could underpin high-capacity coding, all-optical modifying, and buffer of bit signals.

## 4. Conclusion

In summary, we provide the full theoretical and numerical characterization of SCSs in passive Kerr resonators operating in the multi-stable regime. We predicted the coexistence of high-order SCSs, and for the first time, we identified the parameter space of CSs and SCSs based on the CLLEs. Moreover, their stability analysis in the function of the pump power and the cavity detuning uncover the complexity of stable, breathing/chaotic SCSs. All the theoretical predications, including the boundary of the existing map, as well as the stability of CSs and SCSs have been confirmed with the Ikeda-map-based numerical simulations.

Thanks to the accurate parameter space, we may identify the coexistence area of CSs and distinct families of SCSs for the given pump power and the cavity detuning in a fiber ring resonator with moderate experimental parameters. For the first time, we proposed the efficient excitation and manipulation of SCSs by adding the phase/intensity perturbations. This deterministic switching technique of multi-level solitons shows great potential about high-capacity all-optical coding and buffer in passive resonators.

Intuitively, by increasing the pump power, more resonances could be superposed with each other for a given value of cavity detuning. In this scenario, one may expect to excite higher-order of super cavity solitons, which could probably provide more degrees of freedom aiming at more efficient bit coding and buffering based on multi-level cavity solitons. However, we should also mention that for higher orders of SCSs, the pulse width could be of 100s' of fs and the peak power could attend to 100s' W. In this case, the spontaneous Raman scattering (SRS) may play an un-negligible role [41], which may shrink the parameter space. Therefore, different orders of SCSs are prevented from coexisting. Moreover, the SRS might also induce the drifting velocity of cavity solitons, which could limit the bandwidth of the all-optical buffering of the passive fiber resonators. Nevertheless, for broader perspectives about the versatile light source for telecommunications or metrology, the extra degree of freedom for CSs and SCSs could be introduced by considering the polarization components or the multiplexing of their transverse modes. These ambitious research topics involving more complex coupling dynamics and the impact of high-order nonlinearities, either in passive resonators or in mode-locked lasers require further investigations.

## Appendix

In the appendix section, we give the concrete form of the characteristic equation and its stability analysis. we take the complex amplitude of the steady CSs/SCSs from the Ikeda-map as the approximate initial guess, and then we use Newton-Raphson algorithm to converge it to the stationary soliton solution. In 3-CLLEs, we divide three amplitude elements in the complex amplitude of the model into real and imaginary parts to obtain scalar coordinate variables $X$ (e.g., $U_0 = A_0 + iA_1$). As we mentioned, with the rough guessing of the coordinate $X_1$, one may obtain the linear approximation of the function $F(X)$ to approach the desired accurate solution $X_0$. Here, we use the Taylor series expansion of initial guessing $X_1$:

$$\begin{cases} F(X_0) = 0 \\ F(X_0) \approx F(X_1) + J(X_1)(X_0 - X_1) \end{cases} \quad (A1)$$

where $J(X_1)$ represents the Jacobian matrix of $F(X)$ at $X_1$. In fact, an estimate of $X_0$ can be obtained by truncating Eq. (A1) to the first order in the Taylor expansion, resulting in the following iterative process:

$$X_2 = -J(X_1)^{-1}F(X_1) + X_1 \quad (A2)$$

In this way, $X_2, X_3 \ldots X_n$ could be obtained to approach the desired solution $X_0$. In this work, the tolerable discrepancy between the left and the right-hand side of Eq. (3) is set to be $10^{-10}$, which is a trade-off between accuracy and computing efficiency. We should also mention that it is crucial to select $X_1$ to ensure that the Jacobian at $X_1$ is sufficiently close to that at $X_0$, otherwise the iteration process in Eq. (A2) may fail to converge. Next, we focus on the numerical extension of $F(X, \Delta)$ with the continuation calculation by considering the increment/decrement of the cavity detuning. In this step, in order to track all the equilibrium solutions even the unstable one, the pseudo-arc length method is essentially involved to overcome the saddle node bifurcation problem [35].

For the 3-CLLE model, one may divide the amplitudes of the three different modes into real and imaginary parts to get $X$ =$[A_1\ A_2\ A_3\ A_4\ A_5\ A_6]$. In this case, $X$ should satisfy the equations

$$F = F_u + XD_2 \quad (A3)$$

where $D_2$ is the dispersion matrix and $F_u = [f_1\ f_2\ f_3\ f_4\ f_5\ f_6]$ represents other components of the equation $F$. In this case, $D_2$ is with the form of a Kronecker product:

$$D_2 = (C \otimes B)_{6n \times 6n} \quad (A4)$$

Here, the $n \times n$ matrix B and the $6 \times 6$ matrix C could be written as:

$$B = -\frac{\beta_2}{2(dt)^2}\begin{pmatrix} -2 & 1 & \cdots & 0 & 0 \\ 1 & -2 & & 0 & 0 \\ \vdots & & \ddots & & \vdots \\ 0 & 0 & \cdots & -2 & 1 \\ 0 & 0 & & 1 & -2 \end{pmatrix}_{n\times n} \quad (A5)$$

$$C = \begin{pmatrix} 0 & 1 & 0 & 0 & 0 & 0 \\ -1 & 0 & 0 & 0 & 0 & 0 \\ 0 & 0 & 0 & 1 & 0 & 0 \\ 0 & 0 & -1 & 0 & 0 & 0 \\ 0 & 0 & 0 & 0 & 0 & 1 \\ 0 & 0 & 0 & 0 & -1 & 0 \end{pmatrix} \quad (A6)$$

In Eq.(A4-A6), $dt$ and $n$ are the intervals and number of coordinate grids at fast times, respectively. The 1st term on the RHS of Eq. (A3) could be written as:

$$F_u = [f_1\ f_2\ f_3\ f_4\ f_5\ f_6]_{1\times 6n}$$

The 6 elements in the bracket are

$$\begin{aligned}
f_1 &= -\frac{\alpha}{L}(A_1 + A_3 + A_5) - \gamma A_2\big((A_1^2 + A_2^2) + 2(A_3^2 + A_4^2) + 2(A_5^2 + A_6^2)\big) \\
&\quad + \gamma(-2A_3A_4A_5 + A_3^2A_6 - A_4^2A_6) + \frac{\Delta A_2}{L} + \frac{\sqrt{\theta}S}{L} \\
f_2 &= -\frac{\alpha}{L}(A_2 + A_4 + A_6) + \gamma A_1\big((A_1^2 + A_2^2) + 2(A_3^2 + A_4^2) + 2(A_5^2 + A_6^2)\big) \\
&\quad + \gamma(A_3^2A_5 - A_4^2A_5 + 2A_3A_4A_6) - \frac{\Delta A_1}{L} \\
f_3 &= -\frac{\alpha}{L}(A_1 + A_3 + A_5) - \gamma A_4\big(2(A_1^2 + A_2^2) + (A_3^2 + A_4^2) + 2(A_5^2 + A_6^2)\big) \\
&\quad + \gamma(-2A_2A_3A_5 + 2A_1A_4A_5 - 2A_1A_3A_6 - 2A_2A_4A_6) + \frac{(\Delta + 2\pi)A_4}{L} + \frac{\sqrt{\theta}S}{L} \\
f_4 &= -\frac{\alpha}{L}(A_2 + A_4 + A_6) - \gamma A_3\big(2(A_1^2 + A_2^2) + (A_3^2 + A_4^2) + 2(A_5^2 + A_6^2)\big) \\
&\quad + \gamma(2A_1A_3A_5 + 2A_2A_4A_5 - 2A_2A_3A_6 + 2A_1A_4A_6) - \frac{(\Delta + 2\pi)A_3}{L} \\
f_5 &= -\frac{\alpha}{L}(A_1 + A_3 + A_5) - \gamma A_6\big(2(A_1^2 + A_2^2) + 2(A_3^2 + A_4^2) + (A_5^2 + A_6^2)\big) \\
&\quad + \gamma(-2A_1A_3A_4 + A_2A_3^2 - A_2A_4^2) + \frac{(\Delta + 4\pi)A_6}{L} + \frac{\sqrt{\theta}S}{L} \\
f_6 &= -\frac{\alpha}{L}(A_2 + A_4 + A_6) + \gamma A_5\big(2(A_1^2 + A_2^2) + 2(A_3^2 + A_4^2) + (A_5^2 + A_6^2)\big) \\
&\quad + \gamma(A_1A_3^2 - A_1A_4^2 + 2A_2A_3A_4) - \frac{(\Delta + 4\pi)A_5}{L}.
\end{aligned} \quad (A7)$$

Next, by separating the complex amplitude, we can transform the 3-CLLE model into a form of pure real numbers, thus enabling implementation into the Newton Raphson calculation. In this way, one may find the feasible routine to test if the system attains the stable states for the given parameters. To this aim, finding the equilibrium position of the equation $F$ is a necessary procedure. It requires constant updates for the Jacobian matrix with this form:

$$J = \begin{pmatrix} \frac{\partial f_1}{\partial A_1}I & \cdots & \frac{\partial f_1}{\partial A_6}I \\ \vdots & \ddots & \vdots \\ \frac{\partial f_6}{\partial A_1}I & \cdots & \frac{\partial f_6}{\partial A_6}I \end{pmatrix}_{6n\times 6n} + D_2 \quad (A8)$$

In this equation, $I$ is the unit matrix. By means of numerical continuation, we could find the stationary solution of the system by analyzing the eigenvalues of the matrix $J$. For the given coordinate $X$ of a stationary solution, the $\lambda_{6n}$ represents its associated eigenvalues. The possible cases depend on their real and the imaginary parts:

If Re($\lambda_{6n}$) < 0, the disturbance to the system will decay exponentially, indicating that the solution is stable.
If Re($\lambda_{6n}$) > 0, the system is unstable and the perturbations increase exponentially.
If Re($\lambda_{6n}$) = 0, the solitons are at a bifurcation point. In this case, two kinds of bifurcation can occur. There is a Hopf bifurcation when Im($\lambda_{6n}$) ≠ 0, and a saddle node bifurcation when Im($\lambda_{6n}$) = 0.

## Acknowledgements

This work is partially supported by the National Natural Science Foundation of China (62275097), the key developping project of Hubei Province in China (2023BAB062), the Open Project Program of Wuhan National Laboratory for Optoelectronics (2023WNLOKF007) and the Guangdong Basic and Applied Basic Research Foundation (2023A1515010965, 2024A1515010017). The authors are grateful to Massimo Giudici for the precious comments during the PIERS conference held in Chengdu in April 2024.

## References

[1]. Soliton-driven photonics[M]. Springer Science & Business Media, 2012.
[2]. Tikan A, Riemensberger J, Komagata K, et al. Emergent nonlinear phenomena in a driven dissipative photonic dimer[J]. Nature Physics, 2021, 17(5): 604-610.
[3]. Weiner A M. Cavity solitons come of age[J]. Nature Photonics, 2017, 11(9): 533-535.
[4]. Haelterman M, Trillo S, Wabnitz S. Dissipative modulation instability in a nonlinear dispersive ring cavity[J]. Optics communications, 1992, 91(5-6): 401-407.
[5]. Ankiewicz A, Pask C. Chaos in optics: field fluctuations for a nonlinear optical fibre loop closed by a coupler[J]. The ANZIAM Journal, 1987, 29(1): 1-20.
[6]. Leo F, Gelens L, Emplit P, et al. Dynamics of one-dimensional Kerr cavity solitons[J]. Optics express, 2013, 21(7): 9180-9191.
[7]. Coulibaly S, Taki M, Bendahmane A, et al. Turbulence-induced rogue waves in Kerr resonators[J]. Physical Review X, 2019, 9(1): 011054.
[8]. Lugiato L, Prati F, Brambilla M. Nonlinear optical systems[M]. Cambridge University Press, 2015.
[9]. Pasquazi A, Peccianti M, Razzari L, et al. Micro-combs: A novel generation of optical sources[J]. Physics Reports, 2018, 729: 1-81.
[10]. Ikeda K. Multiple-valued stationary state and its instability of the transmitted light by a ring cavity system[J]. Optics communications, 1979, 30(2): 257-261.
[11]. Lugiato L A, Lefever R. Spatial dissipative structures in passive optical systems[J]. Physical review letters, 1987, 58(21): 2209.
[12]. Coen S, Tlidi M, Emplit P, et al. Convection versus dispersion in optical bistability[J]. Physical review letters, 1999, 83(12): 2328.
[13]. Descalzi O, Clerc M, Residori S, et al. (2011). Localized states in physics: solitons and patterns[M]. Springer Science & Business Media, 2011.
[14]. Obrzud. E, S. Lecomte and Herr T. "Temporal solitons in microresonators driven by optical pulses." Nature Photonics 11.9 (2017): 600-607.
[15]. Zheng H, Sun W, Ding X, et al. Programmable access to microresonator solitons with modulational sideband heating[J]. APL Photonics, 2023, 8(12).
[16]. Agrawal G P. Modulation instability induced by cross-phase modulation[J]. Physical review letters, 1987, 59(8): 880.
[17]. Wabnitz S. Modulational polarization instability of light in a nonlinear birefringent dispersive medium[J]. Physical review A, 1988, 38(4): 2018.
[18]. Xia R, et al. "Coupling dynamics of dissipative localized structures: From polarized vector solitons to soliton molecules." Optics Communications (2023): 129996.
[19]. Stratmann M, Pagel T, Mitschke F. Experimental observation of temporal soliton molecules[J]. Physical review letters, 2005, 95(14): 143902.
[20]. Xu G, Nielsen A U, Garbin B, et al. Spontaneous symmetry breaking of dissipative optical solitons in a two-component Kerr resonator[J]. Nature Communications, 2021, 12(1): 4023.
[21]. Xu G, Hill L, Fatome J, et al. Breathing dynamics of symmetry-broken temporal cavity solitons in Kerr ring resonators[J]. Optics Letters, 2022, 47(6): 1486-1489.
[22]. Hansson T, Wabnitz S. Frequency comb generation beyond the Lugiato–Lefever equation: multi-stability and super cavity solitons[J]. JOSA B, 2015, 32(7): 1259-1266.
[23]. Huang T, Zheng H, Xu G, et al. Coexistence of nonlinear states with different polarizations in a Kerr resonator[J]. Physical Review A, 2024, 109(1): 013503.
[24]. Nielsen A U, Garbin B, Coen S, et al. Coexistence and interactions between nonlinear states with different polarizations

in a monochromatically driven passive Kerr resonator[J]. Physical review letters, 2019, 123(1): 013902.

[25]. Anderson M, Wang Y, Leo F, et al. Coexistence of multiple nonlinear states in a tristable passive Kerr resonator[J]. Physical Review X, 2017, 7(3): 031031.

[26]. Conforti M, Biancalana F. Multi-resonant lugiato–lefever model[J]. Optics Letters, 2017, 42(18): 3666-3669.

[27]. Kartashov Y V, Alexander O, Skryabin D V. Multistability and coexisting soliton combs in ring resonators: the Lugiato-Lefever approach[J]. Optics Express, 2017, 25(10): 11550-11555.

[28]. Coen S, Randle H G, Sylvestre T, et al. Modeling of octave-spanning Kerr frequency combs using a generalized mean-field Lugiato–Lefever model[J]. Optics letters, 2013, 38(1): 37-39.

[29]. Szöke A, Daneu V, Goldhar J, et al. Bistable optical element and its applications[J]. Applied Physics Letters, 1969, 15(11): 376-379.

[30]. Coen S, Erkintalo M. Universal scaling laws of Kerr frequency combs[J]. Optics letters, 2013, 38(11): 1790-1792.

[31]. Chembo Y K, Yu N. Modal expansion approach to optical-frequency-comb generation with monolithic whispering-gallery-mode resonators[J]. Physical Review A, 2010, 82(3): 033801.

[32]. Matsko A B, Savchenkov A A, Strekalov D, et al. Optical hyperparametric oscillations in a whispering-gallery-mode resonator: Threshold and phase diffusion[J]. Physical Review A, 2005, 71(3): 033804.

[33]. Arabí C M, Englebert N, Parra-Rivas P, et al. Mode-locking induced by coherent driving in fiber lasers[J]. Optics Letters, 2022, 47(14): 3527-3530.

[34]. Parra-Rivas P, Gomila D, Gelens L, et al. Bifurcation structure of localized states in the Lugiato-Lefever equation with anomalous dispersion[J]. Physical Review E, 2018, 97(4): 042204.

[35]. Krauskopf B, Osinga H M, Galán-Vioque J. Numerical continuation methods for dynamical systems[M]. Berlin: Springer, 2007.

[36]. Barland S, Tredicce J R, Brambilla M, et al. Cavity solitons as pixels in semiconductor microcavities[J]. Nature, 2002, 419(6908): 699-702.

[37]. Firth W J, Scroggie A J. Optical bullet holes: robust controllable localized states of a nonlinear cavity[J]. Physical review letters, 1996, 76(10): 1623.

[38]. Jang J K, Erkintalo M, Coen S, et al. Temporal tweezing of light through the trap and manipulation of temporal cavity solitons[J]. Nature communications, 2015, 6(1): 7370.

[39]. Parra-Rivas P, Hetzel S, Kartashov Y V, et al. Quartic Kerr cavity combs: bright and dark solitons[J]. Optics Letters, 2022, 47(10): 2438-2441.

[40]. Parra-Rivas P, Gomila D, Matías M A, et al. Dynamics of localized and patterned structures in the Lugiato-Lefever equation determine the stability and shape of optical frequency combs[J]. Physical Review A, 2014, 89(4): 043813.

[41]. Wang Y, Anderson M, Coen S, et al. Stimulated Raman scattering imposes fundamental limits to the duration and bandwidth of temporal cavity solitons[J]. Physical Review Letters, 2018, 120(5): 053902.